\documentclass[9pt,twocolumn,twoside]{osajnl}
\usepackage{scalerel}
\usepackage{tikz}
\usetikzlibrary{svg.path,calc}

\definecolor{orcidlogocol}{HTML}{A6CE39}
\tikzset{
   orcidlogo/.pic={
    \fill[orcidlogocol] svg{M256,128c0,70.7-57.3,128-128,128C57.3,256,0,198.7,0,128C0,57.3,57.3,0,128,0C198.7,0,256,57.3,256,128z};
    \fill[white] svg{M86.3,186.2H70.9V79.1h15.4v48.4V186.2z}
                 svg{M108.9,79.1h41.6c39.6,0,57,28.3,57,53.6c0,27.5-21.5,53.6-56.8,53.6h-41.8V79.1z M124.3,172.4h24.5c34.9,0,42.9-26.5,42.9-39.7c0-21.5-13.7-39.7-43.7-39.7h-23.7V172.4z}
                 svg{M88.7,56.8c0,5.5-4.5,10.1-10.1,10.1c-5.6,0-10.1-4.6-10.1-10.1c0-5.6,4.5-10.1,10.1-10.1C84.2,46.7,88.7,51.3,88.7,56.8z};
  }
}

\newcommand\orcidicon[1]{\href{https://orcid.org/#1}{\mbox{
\begin{tikzpicture}[overlay,remember picture]
\coordinate (A);
\coordinate(B) at ($(A)-(2pt,-9pt)$);
\end{tikzpicture}
\begin{tikzpicture}[overlay,remember picture,yscale=-0.0381,xscale=0.0381,transform shape]
\pic at (B) {orcidlogo};
\end{tikzpicture}
}{}}}

\usepackage{hyperref}
 
\journal{ol} 
\setboolean{shortarticle}{true} 

\title{Ultra-low-noise supercontinuum generation with a flat near-zero normal dispersion fiber}
\author[1,*,\protect\orcidicon{0000-0003-2007-0930}\,\,]{ Shreesha Rao D. S.}
\author[1]{Rasmus D. Engelsholm}
\author[1,\protect\orcidicon{0000-0002-2618-0631}\,\,]{Iv\'{a}n B. Gonzalo}
\author[1,\protect\orcidicon{0000-0002-4528-5447}\,\,]{Binbin Zhou}
\author[2]{Patrick~Bowen}
\author[2]{Peter M. Moselund}
\author[1,2]{Ole Bang}
\author[1,\protect\orcidicon{0000-0002-9407-9236}\,\,]{Morten Bache}

\affil[1]{DTU Fotonik, Department of Photonics Engineering, Technical University of Denmark, \O rsteds Plads, 2800 Kongens Lyngby, Denmark.}
\affil[2]{NKT Photonics A/S, Blokken 84, 3460 Birker\o d, Denmark.}

\affil[*]{Corresponding author: shdes@fotonik.dtu.dk}

\begin{abstract}
A pure silica photonic crystal fiber with a group velocity dispersion ($\beta_2$) of 4 ps$^2$/km at 1.55 $\mu$m and less than 7 ps$^2$/km from 1.32 $\mu$m to the zero dispersion wavelength (ZDW) 1.80 $\mu$m was designed and fabricated. The dispersion of the fiber was measured experimentally and found to agree with the fiber design, which also provides low loss below 1.83 $\mu$m due to eight outer rings with increased hole diameter. The fiber was pumped with a 1.55 $\mu$m, 125 fs laser and, at the maximum in-coupled peak power (P$_0$) of 9 kW, a 1.34$-$1.82 $\mu$m low-noise spectrum with a relative intensity noise below 2.2\% was measured. The numerical modeling agreed very well with the experiments and showed that P$_0$ could be increased to 26 kW before noise from solitons above the ZDW started to influence the spectrum by pushing high-noise dispersive waves through the spectrum. 
\end{abstract}

\begin{document}
\maketitle 
Supercontinuum generation (SCG) in fibers by anomalous dispersion pumping (ADP) has been widely studied to obtain a broad spectrum~\cite{dudley02C}. One of the initial studies used a 7.6 ps input pulse to obtain a broadband source for wavelength division multiplexing-based telecommunication~\cite{MorWDM}. ADP with picosecond (ps) and nanosecond lasers has become the method of choice for high average power commercial sources. However, it is well known that SCG with ps or longer pulses is very noisy when pumped both in the anomalous dispersion regime (ADR)~\cite{MolADNoi12} and in the normal dispersion regime (NDR)~\cite{MolNDNoi13}. In contrast, it was observed that SCG by ADP with femtosecond (fs) pulses with soliton number N < 10 leads to reduced noise~\cite{dudley02C}. Another way of generating a coherent SC is by using an all normal dispersion (ANDi) fiber pumped by a fs pulse~\cite{Finot2008}, which can generate an octave spanning SC,  as demonstrated numerically~\cite{HeidtFlatTop} and experimentally~\cite{HeidtOcta} by Heidt \textit{et al}. An ANDi fiber-based SC can result in a smooth spectral intensity and phase profile and preserve a single pulse in the time domain. ANDi fiber-based SC can thus provide excellent coherence, even when pumped by a pulse duration of several hundreds of fs, provided a polarization maintaining (PM) fiber is used~\cite{heidt17L,IvnRIN}. If a PM fiber is not used, the pump pulse length has to be shorter and/or P$_0$ lower to maintain coherence~\cite{IvnRIN}. The good coherence has been shown experimentally by performing multiple single-shot measurements based on a dispersive Fourier transform approach~\cite{KlimDFT}. Spectra generated by these ANDi fibers enable applications that need flat and coherent SCs such as single-beam coherent anti-Stokes Raman spectroscopy (CARS)~\cite{LiuCars}, pulse compression~\cite{heidt2011}, optical coherence tomography (OCT)~\cite{maria2017q}, multimodal nonlinear microscopy~\cite{LiuMul}, ultrafast transient spectroscopy~\cite{HeidtOcta}, multi-spectral imaging~\cite{CRP18Mid}, and telecommunications~\cite{MorWDM}.

Although SCG in flat and close to zero ANDi fibers has been extensively studied, particularly pumped around 1 $\mu$m, there are no commercially available ANDi fibers at 1.55 $\mu$m. Our aim is to design such an ANDi fiber that allows us to obtain a flat and coherent SC using a low P$_0$ table top laser with a reasonably long pulse length ($\sim$120 fs) at 1.55 $\mu$m. Several attempts have been made to design an ANDi fiber with a flat dispersion around 1.55 $\mu$m~\cite{Sai04FibDes,huang2018A,Huang2018H,Tar2016all,Hu2008,Hai08Uni,hartung2011design,Olyaee12Ult,chat2015des,Suk2017des,Fer2018Des}, but only two fibers have been fabricated and demonstrated to support SCG. One was an all-solid photonic crystal fiber (PCF) made of soft glasses, which had a $\beta_2$ below 30 ps$^2$/km~\cite{KlimOcta}. Another was a silica PCF doped with germanium~\cite{tarnowski2016} to provide a $\beta_2$ within 0 to 10 ps$^2$/km.

In this Letter, we have designed and drawn a PCF made of pure silica with a hole structure similar to the one proposed in Ref.~\cite{hartung2011design}. In contrast to Ref.~\cite{hartung2011design}, we use a non-uniform hole diameter design with 11 rings, where rings 4 to 11 have an increased diameter in order to improve confinement and reduce the loss. Thus, our silica PCF has a special hole structure to give the desired low dispersion without the need of doping or using non-standard glasses. In this Letter, we report the fabrication of such a fiber, along with experimental measurement of its dispersion and experimental and numerical studies of the generated SC and its noise properties. 

The main factors considered for the fiber design with respect to optimal fs pumped SCG are as follows. (1) \textit{Spectral broadening}: in an ANDi fiber, the spectrum broadens coherently by self-phase modulation (SPM) and optical wave breaking (OWB)~\cite{Finot2008}. The bandwidth is independent of pulse length and given by $\Delta \omega_{coh} \propto (\gamma P_0/\beta_2)^{1/2}$, where $\gamma$ is the nonlinear coefficient~\cite{heidt17L}. To obtain a broader spectrum with constant $\gamma$ and P$_0$, we need a lower $\beta_2$. When the dispersion profile of the fiber is asymmetric with respect to the pump, a preferential broadening towards the wavelength region with low values of dispersion can be observed~\cite{HeidtOcta}. (2) \textit{Parametric Raman (PR) amplified noise}: the limiting factor for coherence of the SC in a PM-ANDi fiber is the PR amplified noise~\cite{heidt17L,RamIvn18}. The PR amplified noise becomes important when the PR gain length, L$^{\star}_R$ $(=1/g_s^\star P_0)$, is smaller than the characteristic length of broadening from OWB (L$_{WB}\propto$T$_0[\gamma P_0 \beta_2]^{-1/2}$, where $T_0$ is the intensity 1/e width of a Gaussian). The PR gain, $g_s^\star=2\gamma Re\big\{[K(2q-K)]^{1/2}\big\}$, where K$=-\Delta\beta/2\gamma P_0$, and q$=(1-f_R)+f_R\widetilde{\chi}^{(3)}_R(-\Omega)$, where $\Delta\beta=\beta_2\Omega^2+\frac{\beta_4\Omega^4}{12}+\ldots$, $\widetilde{\chi}^{(3)}_R(-\Omega)$ is the complex Raman susceptibility, and $\Omega$ is the frequency shift with respect to the pump. It can be seen from the expression for $g_s^\star$ that weak and flat dispersion ($\beta_{2,4}$ small) has the advantage that the PR gain is small. This makes the PM fiber suitable to generate coherent SC, even when pumped with a pulse duration of several hundreds of fs. (3) \textit{Spectral flatness}: for $\beta_2$ close to zero, the spectrum is flat, but a $\beta_2$ too close to zero leads to the depletion of the pump during OWB and results in close to 10 dB dips in the spectrum~\cite{HeidtFlatTop}. (4) \textit{Polarization mode instability (PMI)}: it was shown in Ref.~\cite{IvnRIN} that in a single-mode fiber PMI leads to de-coherence and severely limits the fiber length and T$_0$ below which good coherence is achieved, e.g., from 1 ps to 120 fs. Since the PMI gain depends on P$_0$, a reasonably low power is also required for good coherence~\cite{IvnRIN}. Fiber designs with a low value of $\beta_2$ can generate a broad spectrum with low P$_0$. This would avoid the PMI gain and, thus, limit the noise associated with it. (5) \textit{Fiber loss}: the hole structure in the fiber cannot be arbitrary chosen, since good confinement is necessary for low loss, i.e., too small holes will give loss that is too high.

\begin{figure}[b!]
\centering
\includegraphics[width=0.98\linewidth]{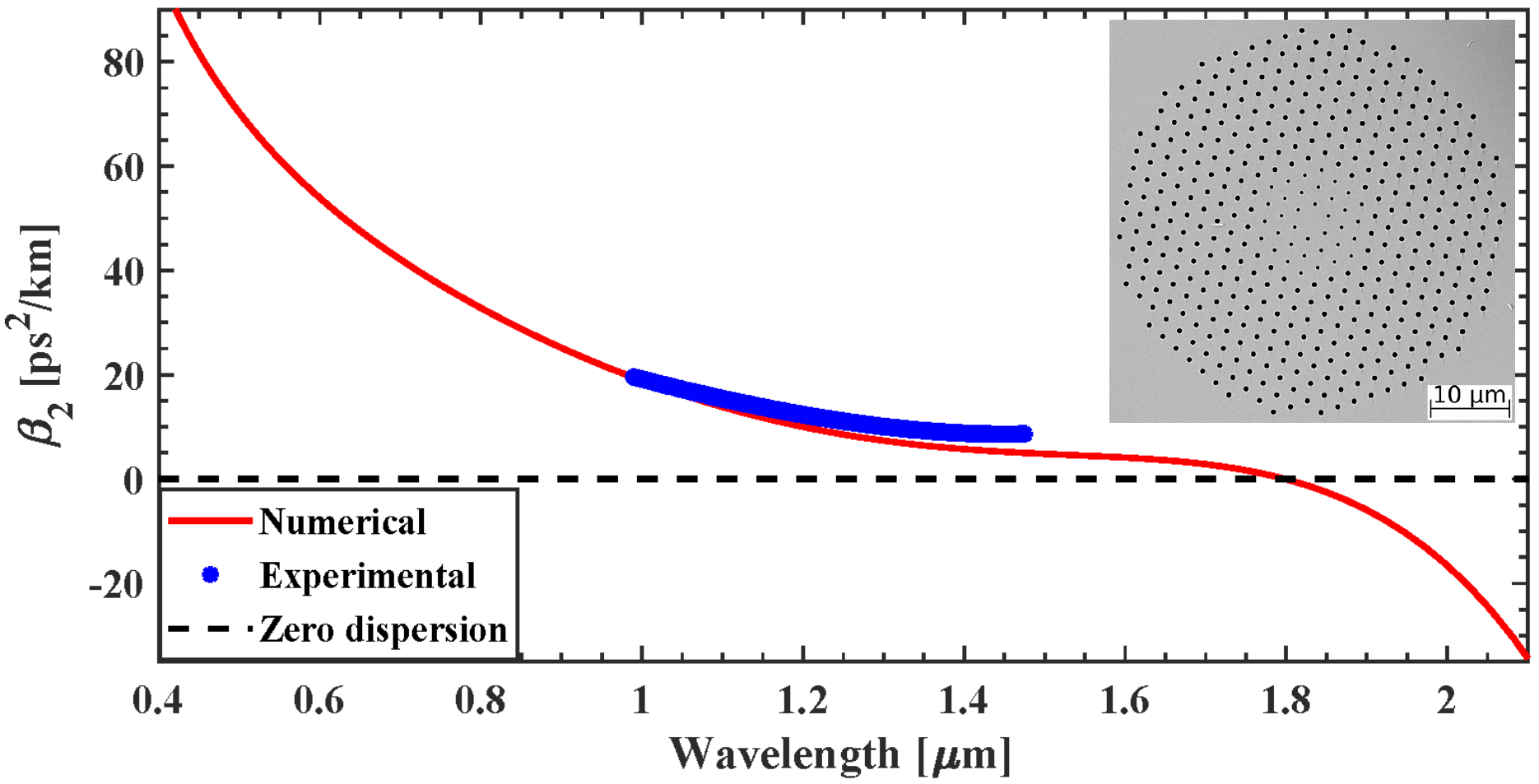}
\caption{Experimentally measured dispersion (dots) of the fiber along with the numerically calculated (solid line) dispersion ($\beta_2$) profile. SEM image of the fiber is in the inset.}
\label{fig:Dis} 
\rule{\linewidth}{1pt}
\end{figure}

The optimized pure silica PCF has a hexagonal hole structure with 11 complete rings with a pitch ($\Lambda$) of 2.3 $\mu$m. The inner three rings of holes have a hole diameter, d$_1\sim$ 550 nm, as is visible from the SEM image in Fig.~\ref{fig:Dis}, which results in flat and normal dispersion. The outer eight rings have a larger hole diameter, d$_2\sim$ 680 nm, in order to reduce the confinement loss in the fiber. 
\begin{figure}[b!]
\centering
\includegraphics[width=0.98\linewidth]{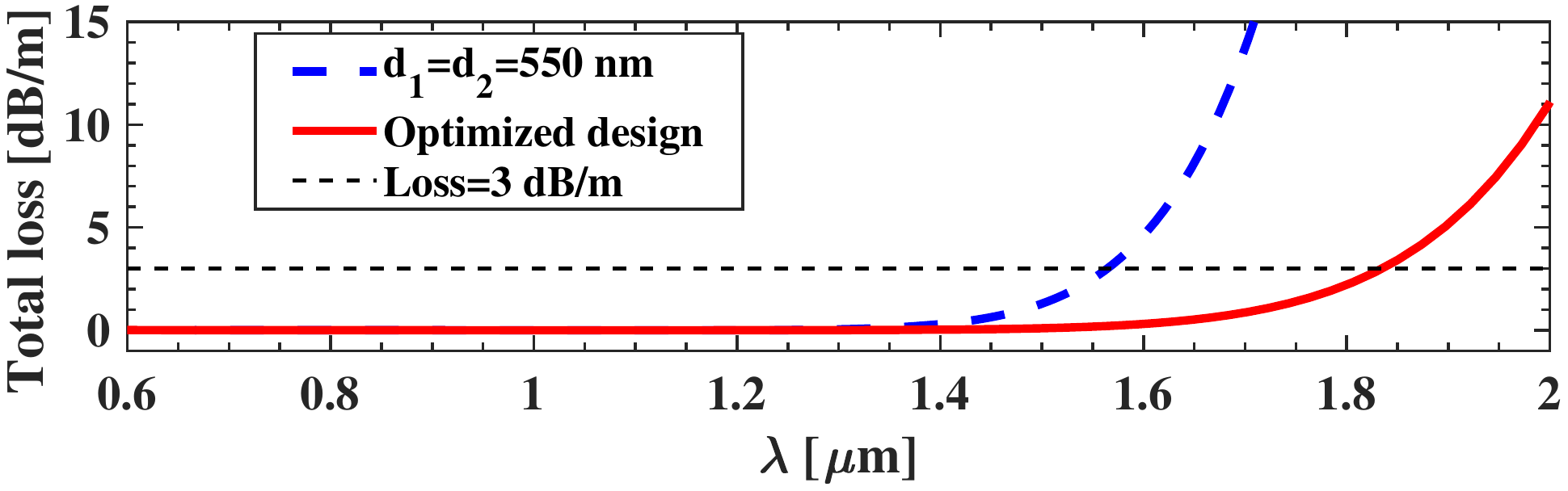}
\caption{Fiber loss in a PCF with $\Lambda=$ 2.3 $\mu$m and d$=$ 550 nm (dashed line) and with the optimized design (solid line).}
\label{fig:Loss} 
\end{figure}
In Fig.~\ref{fig:Loss}, we plot the fiber loss for the optimized PCF (solid line) and for a PCF with a uniform hole diameter of 550 nm (dashed line). The fiber loss is found as the numerically calculated confinement loss, plus the fiber material loss estimated in Ref.~\cite{Mosl09The} (neglecting the imperfection loss). For the design with a uniform hole diameter, the 3 dB/m loss edge is 1.56 $\mu$m. The optimized design with a d$_2$ of 680 nm significantly reduces the confinement loss and pushes the 3 dB/m loss edge to 1.83 $\mu$m.

A full-vectorial finite-element method was used to numerically find the properties of the optimized fiber. The fiber was found to have a $\gamma$ of 3.5 (W$\cdotp$km)$^{-1}$ and an effective area of A$_{eff}=$ 29.9 $\mu$m$^2$ at 1.55 $\mu$m. The dispersion, which is plotted in Fig.~\ref{fig:Dis} (solid line), is indeed found to be flat with a $\beta_2$ of 4 ps$^2$/km at the 1.55 $\mu$m pump and less than 7 ps$^2$/km in the entire wavelength region 1.32$-$1.80 $\mu$m. To check whether the calculated dispersion accurately models the dispersion of the fabricated fiber the dispersion was experimentally measured using white light interferometry~\cite{HluDis12} and is also plotted in Fig.~\ref{fig:Dis} (dots). The two dispersions are found to agree well.

\begin{figure}[b!]
\centering
\includegraphics[width=0.98\linewidth]{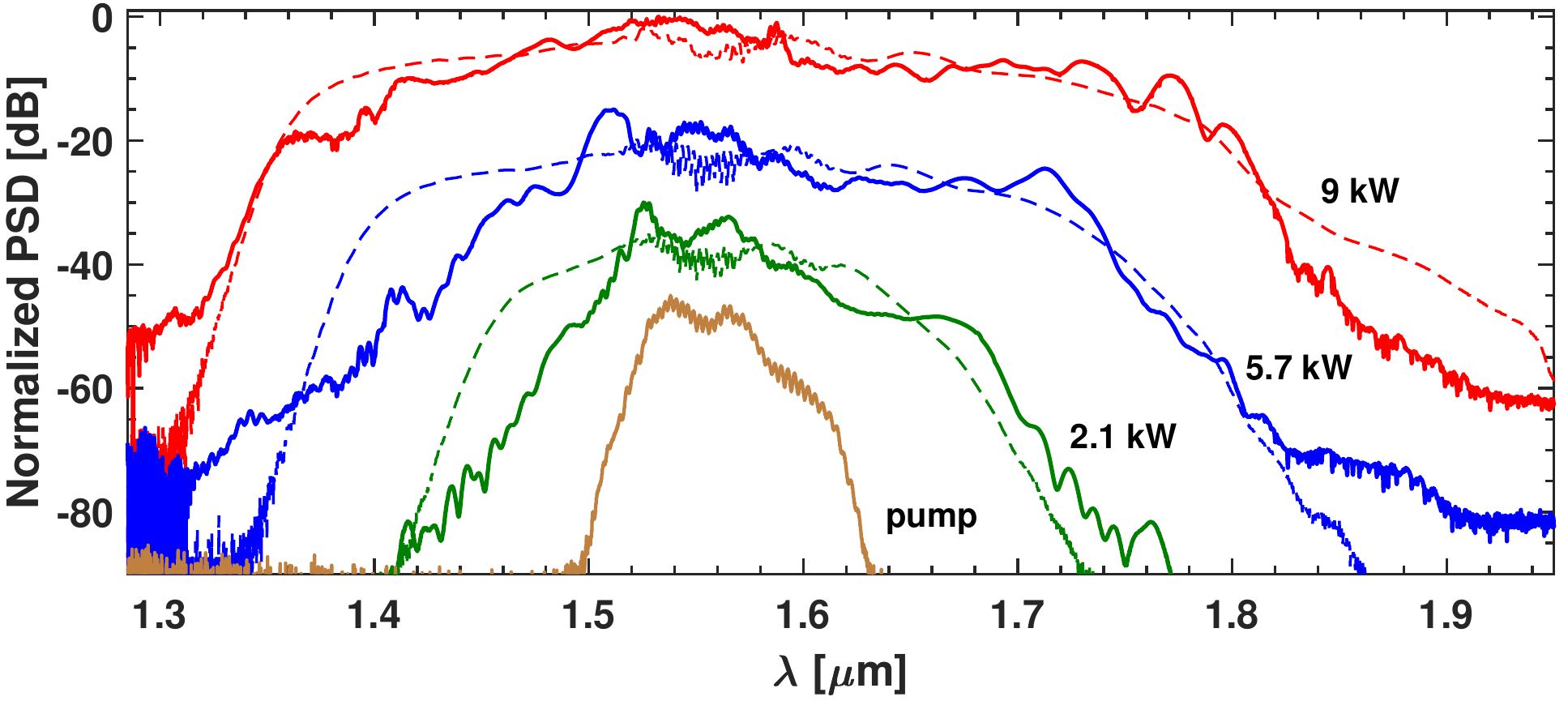}
\caption{Numerical simulations (dashed) and experimentally measured spectra (solid) for different P$_0$ coupled into 3 m of the fiber. A 15 dB offset per curve is provided for clarity.}
\label{fig:VarPeak} 
\end{figure}

SCG in the fiber was investigated by launching a pulse with an intensity full width at half-maximum pulse length (T$_{FWHM}$) of 125 fs at a repetition rate of 90 MHz from a fiber laser (Toptica). Polarized light from the laser was passed through a half-wave plate and coupled into the fiber using an aspheric lens. The coupling efficiency into the fiber, calculated as the ratio of output to input power, taking into account the wavelength dependent loss in the fiber was 53\%, giving a maximum launched P$_0$ of 9 kW and pulse energy of 1.27 nJ. The spectrum was measured using an optical spectrum analyzer (Yokogawa: AQ6375). The measured SC is shown in Fig.~\ref{fig:VarPeak} (solid lines) at different power levels. The $-$30 dB bandwidth of the spectrum covers 1.34$-$1.82 $\mu$m at the maximum power.

The spectral evolution inside the fiber was investigated numerically by solving the single polarization generalized nonlinear Schr\"{o}dinger equation for the envelope function A(z,t) in the interaction picture~\cite{Hult2007}, using the loss and numerically found dispersion. The frequency dependence of $\gamma$ was included through the mode profile dispersion~\cite{Laeg2007Mode}. To have an accurate input pulse for the modelling, we measured the pump power spectral density (PSD) (see Fig.~\ref{fig:VarPeak}) and scaled it appropriately to get the launched spectral intensity, which was then inverse Fourier transformed to obtain the input pulse A(0,t) in the time domain, assuming that the Fourier transform of A(0,t) is real. This gives an input pulse, which is approximately sech shaped with a T$_{FWHM}$ of 125 fs. An independent autocorrelation measurement fitted to a sech confirmed that T$_{FWHM}=$ 125 fs. The numerically found spectra are overlaid with the experimental spectra in Fig.~\ref{fig:VarPeak}. In general, a good correspondence is observed, which means that we can trust the modelling and use it to go to even higher powers.

\begin{figure}[b!]
\centering
\includegraphics[width=0.98\linewidth]{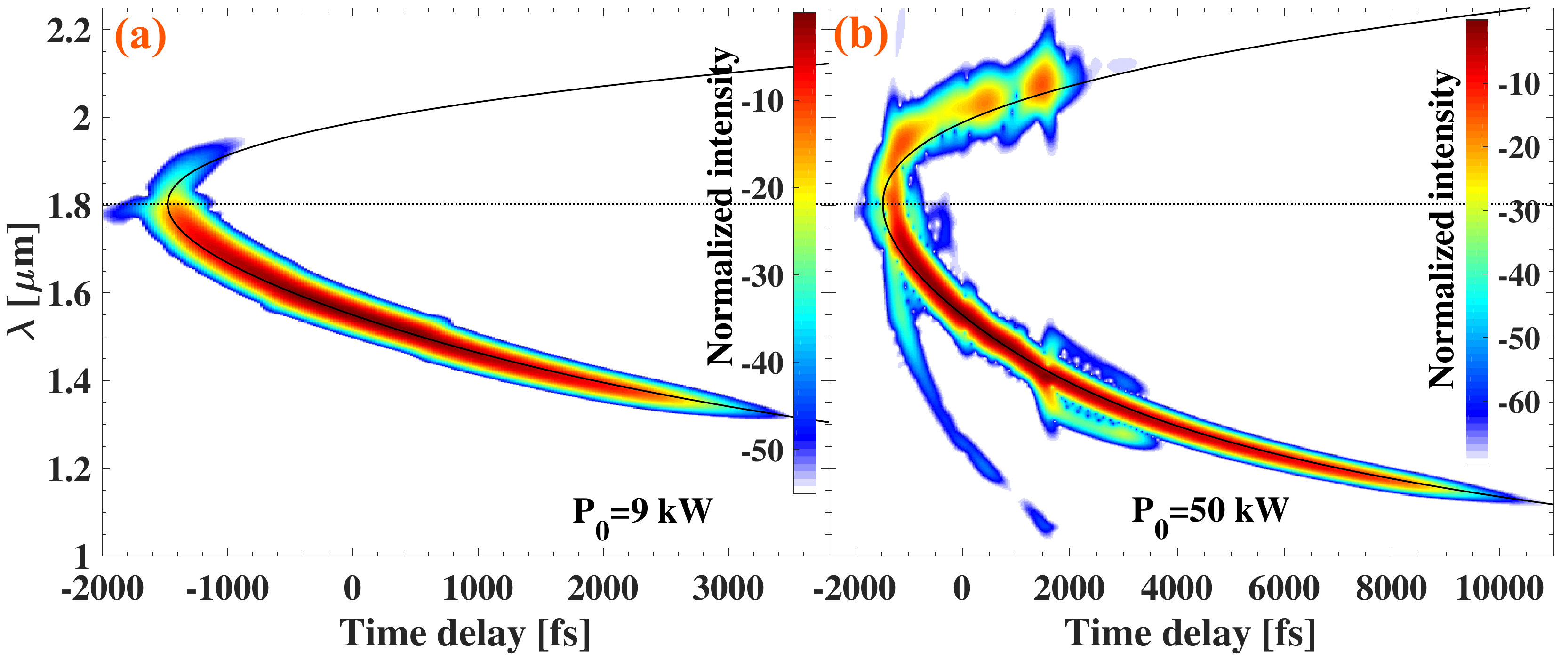}
\caption{Numerically calculated spectrograms at z$=$ 3 m for (a) P$_0=$ 9 kW  and (b) 50 kW . The solid line is the accumulated delay with respect to the pump [z/v$_g$($\lambda$)-z/v$_g$($\lambda_{pump}$)], where v$_g$ is the group velocity, and the dotted line is the ZDW.}
\label{fig:Spectgm} 
\end{figure}

\begin{figure*}[htbp!]
\includegraphics[width=0.98\linewidth]{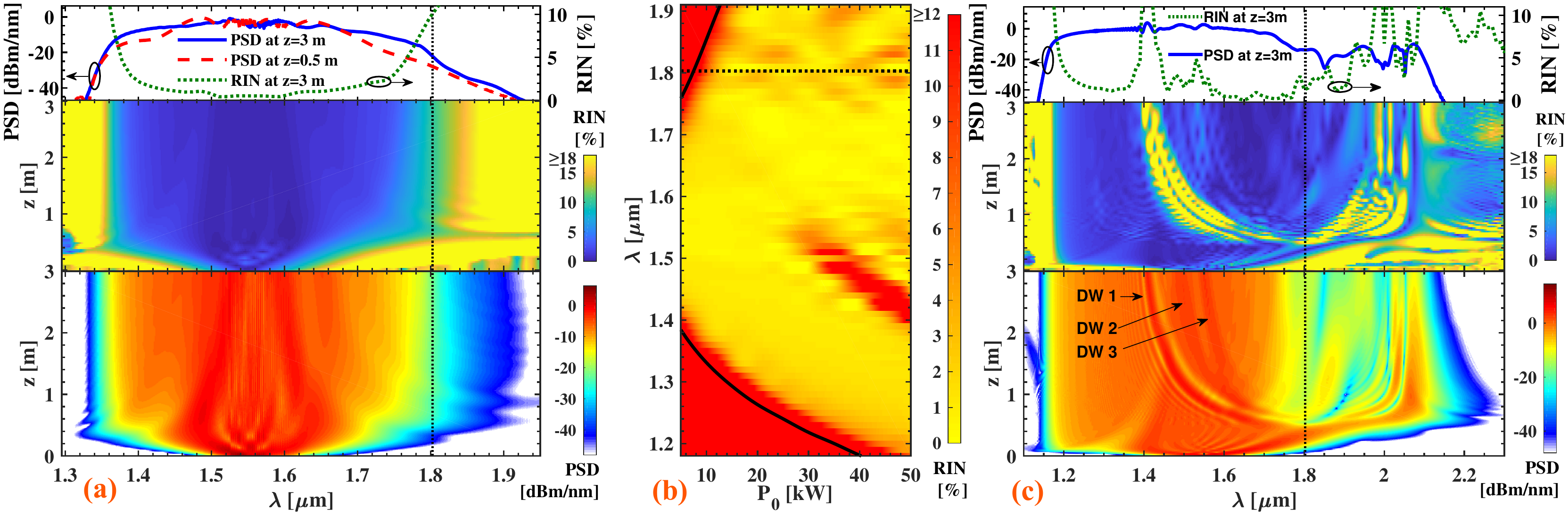}
\caption{Numerical simulations with T$_{FWHM}=$ 125 fs. (a) Top: PSD after 0.5 m (dashed), and 3 m (solid); and RIN after 3m (dotted) for P$_0=$ 9 kW. Below: PSD and RIN evolution along the length of the fiber for P$_0=$ 9 kW. (b) Output RIN at z$=$ 3 m versus P$_0$. The two solid lines are the $-$30 dB spectral edges. (c) Top: PSD (dashed) and RIN (dotted) after 3 m for P$_0=$ 50 kW. Below: PSD and RIN evolution along the length of the fiber for P$_0=$ 50 kW.}
\label{fig:SpecEvol} 
\rule{\linewidth}{1pt}
\end{figure*}

For the maximum power of the laser with P$_0=$ 9 kW and T$_{FWHM}=$ 125 fs we observe that the spectral broadening primarily takes place within the first 0.5 m, and further propagation up to 3 m just flattens the spectrum, as seen in Fig.~\ref{fig:SpecEvol}(a). At this low pump power all power is seen to remain in the NDR below the zero dispersion wavelength (ZDW), and the spectral evolution resembles to that of standard ANDi fiber-based SCG. The spectrogram at the output of the fiber is plotted in Fig.~\ref{fig:Spectgm}(a). It can be observed that the output pulse chirp is approximately linear and, therefore, suitable for external compression. 

\begin{figure}[b!]
\includegraphics[width=0.98\linewidth]{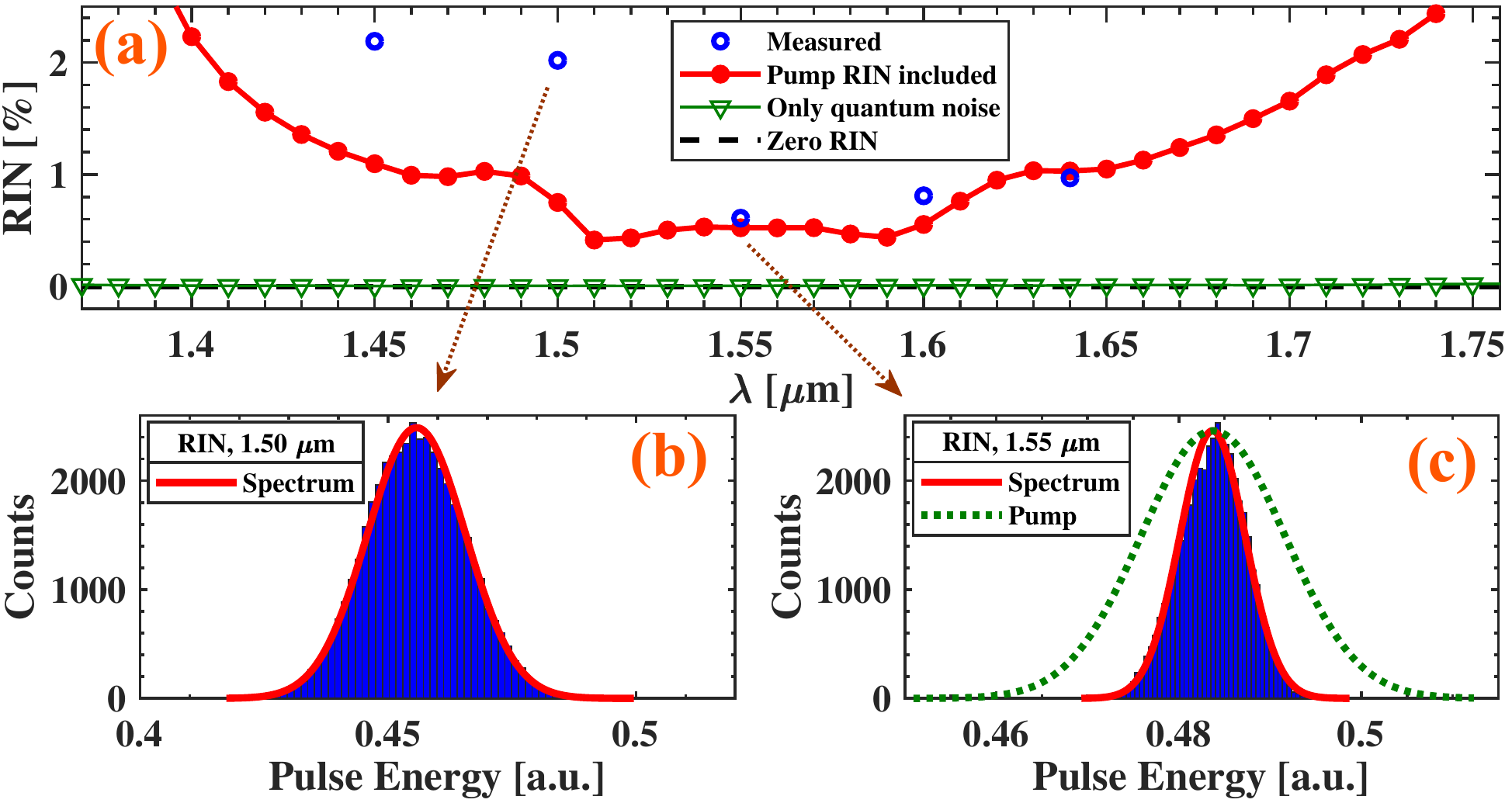}
\caption{(a) Numerically calculated and experimentally measured RIN. (b) Measured histogram for the SC: 12 nm of SC at 1.50 $\mu$m with the fit to Gaussian distribution. (c) Example for RIN measurement at 1.55 $\mu$m: pump (dotted) and SC (solid).}
\label{fig:RIN} 
\end{figure}

To experimentally measure the relative intensity noise (RIN), the spectrum for the maximum power of P$_0=$ 9 kW was filtered using 12 nm bandpass filters and then measured using a large bandwidth (5 GHz) InGaAs detector (Thorlabs: DET08CFC). The voltage versus time trace from the detector were recorded using a large bandwidth (4 GHz bandwidth, 40 GS/s) oscilloscope (Teledyne LeCroy: HDO9404). In all the RIN measurements, the oscilloscope trace was recorded for 0.5 ms corresponding to 45135 pulses. We did not measure the RIN below 1.45 $\mu$m because of too low power (<50 $\mu$W in the 12 nm bandwidth after the filtering optics at 1.4 $\mu$m) and above 1.65 $\mu$m, because the photodiode response tapers off here. The low noise measured supports the fact that the spectrum broadens deterministically by SPM and OWB. The peaks of the measured trace, which are proportional to the pulse energy, were extracted and used to find the RIN$=\sigma/\mu$, where $\sigma$ is the standard deviation, and $\mu$ is the mean. The measured RIN values are shown in Fig.~\ref{fig:RIN}(a). Examples of the approximately Gaussian distribution of the pulse energy at 1.50 and 1.55 $\mu$m are shown in Figs.~\ref{fig:RIN}(b) and~\ref{fig:RIN}(c). 

For the modelling, we also measured the RIN of the pump to be 1\%. Interestingly, the RIN of the SC at 1.55 $\mu$m is smaller than the RIN of the pump, as also seen from the difference in standard deviation in Fig.~\ref{fig:RIN}(c).

For comparison with experiments, we calculated the RIN by adding quantum noise $\delta A(t)$, containing independent normally distributed real and imaginary parts, in each time bin with width $\Delta t$ of the input envelope function in the Wigner representation. The quantum noise has a variance of $\hbar\omega_0/2\Delta t$~\cite{Corney01QN2,ZhBa2016Noi}, where $\omega_0$ is the pump frequency. All RIN calculations were done using an ensemble of 20 independent simulation with different noise seeds. The spectral flatness and extremely low value of the numerically calculated RIN, including only the quantum noise which is plotted in Fig.~\ref{fig:RIN}(a) (green curve), cannot explain the measured noise. The low noise is to be expected from theory, since the PR amplified noise is suppressed by the small value of $\beta_2$, the short T$_{FWHM}$ of 125 fs, the moderate fiber length, and the low P$_0$ of 9 kW~\cite{heidt17L}, which also assures that PMI is avoided~\cite{IvnRIN}. 

The coherent nature of ANDi fiber-based SCG, reflected in the measured low RIN values below 2.2\%, means that the noise of the pump laser suddenly becomes very important, as also briefly mentioned in Ref.~\cite{IvnRIN}, and seen by the fact that we measure an SC noise at 1.55 $\mu$m below that of the pump. Therefore, we re-calculated the RIN of the SC numerically by, in addition the quantum noise, adding 1\% fluctuation in P$_0$ and keeping P$_0$T$_0$ a constant to reflect that in the laser the pulse length decreases when the P$_0$ increases. This gives the red curve in Fig.~\ref{fig:RIN}(a), which clearly better reflects the measured noise. 

To get a broad bandwidth of weak normal dispersion around 1.55 $\mu$m with the chosen fiber design, we had to accept the presence of a ZDW at 1.8 $\mu$m. This means that at a certain power level (above the maximum of our laser), power will cross into the ADR and potentially generate noisy solitons and dispersive waves (DWs)~\cite{Beaud87TDW}, as also investigated for ps-pumped SCG in Ref.~\cite{MolNDNoi13} and fs pulses in Ref.~\cite{RoyRamSo11}. To find this power threshold, we performed numerical simulations with both quantum noise and 1\% laser noise, as detailed above, taking the P$_0$ level to 50 kW.

As an example of SCG at high input P$_0$, the spectral evolution and the output PSD profile for P$_0=$ 50 kW are shown in Fig.~\ref{fig:SpecEvol}(c). For this case, the spectrum initially evolves from SPM, and a part of the pulse crosses over to the ADR at 0.5 m and develops into solitons. These solitons are initially pushed towards longer wavelengths by spectral recoil from what resembles a DW in the NDR. Later, the solitons separate with each of their trapped DWs [marked DW in Fig.~\ref{fig:SpecEvol}(c)] in the NDR, as seen from the spectrogram in Fig.~\ref{fig:Spectgm}(b). Interestingly, the solitons and trapped DWs are seen as high-noise localized spectral features in the evolution of the RIN seen in Fig.~\ref{fig:SpecEvol}(c). The solitons later lose power and, therefore, stop redshifting because of high loss at around 2.1 $\mu$m.  

In Fig.~\ref{fig:SpecEvol}(b), we show the output RIN profile versus pump P$_0$, focussing on the normal dispersion part. We see that the noisy DWs come in and deteriorate the RIN of the SC at around 26 kW, at which the $-$30 dB short wavelength edge has reached 1.24 $\mu$m.

In conclusion, we have presented the design and fabrication of a pure silica PCF with weak normal dispersion in the region 1.32$-$1.8 $\mu$m, which is suitable for low-noise ANDi-based SCG pumped at 1.55 $\mu$m. A two hole size structure was used to keep the loss low below the ZDW of 1.8 $\mu$m and, experimentally, an ultra-low noise 1.34$-$1.8 $\mu$m SC with a RIN below 2.2\% was demonstrated using a 1.55 $\mu$m, 125 fs, 90 MHz fiber laser at the maximum average power of 216 mW (P$_0$ of 9 kW at 53\% in-coupling). The numerical modelling was shown to reproduce the experimental SC and used to show that the power could be increased by about a factor 3 without degrading the noise to extend the low-noise spectrum to 1.24$-$1.8 $\mu$m. A detailed RIN investigation revealed how high-noise localized DWs moved through the NDR trapped by solitons in the ADR when the pump power was too high. The experimentally generated spectrum is suitable for external compression, single-beam CARS, and telecommunications, as well as in optical imaging applications such as 1.7 $\mu$m OCT. 

\paragraph{Funding.} Horizon 2020 Framework Programme Marie Curie grant No. (722380 [SUPUVIR]); Innovationsfonden (4107-00011A); Det Frie Forskningsr\r{a}d (DFF) (LOISE-4184-00532B).
\bibliography{AllNorBib}
\bibliographyfullrefs{AllNorBib}
\end{document}